Solubilization of Proteins in 2D Electrophoresis: An Outline

Thierry Rabilloud

1. Introduction

The solubilization process for 2D electrophoresis has to achieve four parallel goals:

1. Breaking macromolecular interactions in order to yield separate polypeptide chains. This includes denaturing the proteins to break noncovalent interactions, breaking disulfide bonds, and disrupting noncovalent interactions between proteins and non-proteinaceous compounds such as lipids or nucleic acids.

2. Preventing any artefactual modification of the polypeptides in the solubilization medium. Ideally, the perfect solubilization medium should freeze all the extracted polypeptides in their exact state prior to solubilization, both in terms of amino acid composition and in terms of post-translational modifications. This means that all the enzymes able to modify the proteins must be quickly and irreversibly inactivated. Such enzymes include of course proteases, which are the most difficult to inactivate, but also phosphatases, glycosidases etc.  In parallel, the solubilization protocol should not expose the polypeptides to conditions in which chemical modifications (e.g. deamidation of Asn and Gln, cleavage of Asp-Pro bonds) may occur.

3. Allowing the easy removal of substances that may interfere with 2D electrophoresis. In 2D, proteins are the analytes. Thus, anything in the cell but proteins can be considered as an interfering substance. Some cellular compounds (e.g. coenzymes, hormones) are so dilute they go unnoticed. Other compounds (e.g. simple non-reducing sugars) do not interact with proteins or do not interfere with the electrophoretic process. However, many compounds bind to proteins and/or interfere with 2D, and must be eliminated prior to electrophoresis if their amount exceeds a critical interference threshold. Such compounds mainly include salts, lipids, polysaccharides (including cell walls) and nucleic acids.

4. Keeping proteins in solution during the 2D electrophoresis process.

Although solubilization stricto sensu stops at the point where the sample is loaded onto the first dimension gel, its scope can be extended to the 2D process per se, as proteins must be kept soluble till the end of the second dimension. Generally speaking, the second dimension is a SDS gel, and very few problems are encountered once the proteins have entered the SDS PAGE gel. The one main problem is overloading of the major proteins when micropreparative 2D is carried out, and nothing but scaling-up the SDS gel (its thickness and its other dimensions) can counteract overloading a SDS gel. However, severe problems can be encountered in the IEF step. They arise from the fact that IEF must be carried out in low ionic strength conditions and with no manipulation of the polypeptide charge. IEF conditions give problems at three stages:

a. During the initial solubilization of the sample, important interactions between proteins of widely different pI and/or between proteins and interfering compounds (e.g. nucleic acids) may happen. This yields poor solubilization of some components.

b. During the entry of the sample in the focusing gel, there is a stacking effect due to the transition between a liquid phase and a gel phase with a higher friction coefficient. This stacking increases the concentration of proteins and may give rise to precipitation events.

c. At, or very close to, the isoelectric point, the solubility of the proteins comes to a minimum. This can be explained by the fact that the net charge comes close to zero, with a concomitant reduction of the electrostatic repulsion between polypeptides. This can also result in protein precipitation or adsorption to the IEF matrix.

Apart from breaking molecular interactions and solubility in the 2D gel which are common to all samples, the solubilization problems encountered will greatly vary from a sample type to another, due to wide differences in the amount and nature of interfering substances and/or spurious activities (e.g. proteases). The aim of this outline chapter is not to give detailed protocols for various sample types, and the reader should refer to the chapters of this book dedicated to the type of sample of interest. I would rather like to concentrate on the solubilization rationale and to describe nonstandard approaches to solubilization problems. A more detailed review on solubilization of proteins for electrophoretic analyses can be found elsewhere [1].

## 2. Rationale of Solubilization-Breaking Molecular Interactions

Apart from disulfide bridges, the main forces holding proteins together and allowing binding to other compounds are non-covalent interactions. Covalent bonds are encountered mainly between proteins and some coenzymes. The non-covalent interactions are mainly ionic bonds, hydrogen bonds and "hydrophobic interactions". The basis for "hydrophobic interactions" is in fact the presence of water. In this very peculiar (hydrogen-bonded, highly polar) solvent, the exposure of nonpolar groups to the solvent is thermodynamically not favored compared to the grouping of these apolar groups together. Indeed, although the van der Waals forces give an equivalent contribution in both configurations, the other forces (mainly hydrogen bonds) are maximized in the latter configuration and disturbed in the former (solvent destruction). Thus, the energy balance in clearly in favor of the collapse of the apolar groups together [2]. This explains why hexane and water are not miscible, and also that the lateral chain of apolar amino acids (L, V, I, F, W, Y) pack together and form the hydrophobic cores of the proteins [3]. These hydrophobic interactions are also responsible for some protein-protein interactions and for the binding of lipids and other small apolar molecules to proteins.

The constraints for a good solubilization medium for 2D electrophoresis are therefore to be able to break ionic bonds, hydrogen bonds, hydrophobic interactions and disulfide bridges under conditions compatible with IEF, i.e. with very low amounts of salt or other charged compounds (e.g. ionic detergents).

2.1. Disruption of disulfide bridges

Breaking of disulfide bridges is usually achieved by adding to the solubilization medium an excess of a thiol compound. Mercaptoethanol was used in the first 2D protocols [4], but its use does have drawbacks. Indeed, a portion of the mercaptoethanol will ionize at basic pH, enter the basic part of the IEF gel and ruin the pH gradient in its alkaline part because of its buffering power [5]. Although its pK is around 8, dithiothreitol is much less prone to this drawback, as it is used at much lower concentrations (usually 50 mM instead of the 700 mM present in 5% mercaptoethanol). However, DTT is still not the perfect reducing agent. Some proteins of very high cysteine content or with cysteines of very high reactivity are not fully reduced by DTT. In these cases, phosphines are very often an effective answer. First, the reaction is stoichiometric, which allows in turn to use very low concentration of the reducing agent (a few mM). Second, these reagents are not as sensitive as thiols to dissolved oxygen. The most powerful compound is tributylphosphine, which was the first phosphine used for disulfide reduction in biochemistry [6]. However, the reagent is volatile, toxic, has a rather unpleasant odor, and needs an organic solvent to make it water-miscible. In the first uses of the reagent, propanol was used as a carrier solvent at rather high concentrations (50%) [6]. It was however found that DMSO or DMF are suitable carrier solvents, which enable the reduction of proteins by 2mM tributylphosphine [7]. All these drawbacks have disappeared with the introduction of a water-soluble phospine, tris (carboxyethyl) phosphine, for which 1M aqueous stock solutions can be easily prepared and stored frozen in aliquots.

2.2. Disruption of noncovalent interactions

The perfect way to disrupt all types of noncovalent interactions would be the use of a charged compound that disrupts hydrophobic interactions by providing a hydrophobic environment. The hydrophobic residues of the proteins would be dispersed in that environment and not clustered together. This is just the description of SDS, and this explains why SDS has been often used in the first stages of solubilization [8-11]. However, SDS is not compatible with IEF, and must be removed from the proteins during IEF (see below).

The other way of breaking most noncovalent interactions is the use of a chaotrope. It must be kept in mind that all the noncovalent forces keeping molecules together must be taken into account with a comparative view on the solvent. This means that the final energy of interaction depends on the interaction per se and on its effects on the solvent. If the solvent parameters are changed (dielectric constant, hydrogen bond formation, polarizability, etc.), all the resulting energies of interaction will change. Chaotropes, which alter all the solvent parameters, exert profound effects on all types of interactions. For example, by changing the hydrogen bond structure of the solvent, chaotropes disrupt hydrogen bonds but also decrease the energy penalty for exposure of apolar groups and therefore favor the dispersion of hydrophobic molecules and the unfolding of the hydrophobic cores of a protein [12]. Unfolding the proteins will also greatly decrease ionic bonds between proteins, which are very often not very numerous and highly dependent of the correct positioning of the residues. As the gross structure of proteins is driven by hydrogen bonds and hydrophobic interactions, chaotropes decrease dramatically ionic interactions both by altering the dielectric constant of the solvent and by denaturing the proteins, so that the residues will no longer positioned correctly. Nonionic chaotropes, as those used in 2D, however are unable to disrupt ionic bonds when high charge densities are present (e.g. histones, nucleic acids) [13]. In this case, it is often quite advantageous to modify the pH and to take advantage of the fact that the ionizable groups in proteins are weak acids and bases. For example, increasing the pH to 10 or 11 will induce most proteins to behave as anions, so that ionic interactions present at pH 7 or lower turn into electrostatic repulsion between the molecules, thereby promoting solubilization. The use of a high pH results therefore in dramatically improved solubilizations, with yields very close to what is obtained with SDS [14]. The alkaline pH can be obtained either by addition of a few mM of potassium carbonate to the urea-detergent-ampholytes solution [14], or by the use of alkaline ampholytes [11], or by the use of a spermine-DTT buffer which allows better extraction of nuclear proteins [15].

For 2D electrophoresis, the chaotrope of choice is urea. Although urea is less efficient than substituted urea in breaking hydrophobic interactions [12], it is more efficient in breaking hydrogen bonds, so that its overall solubilization power is greater. However, denaturation by urea induces the exposure of the totality of the proteins hydrophobic residues to the solvent. This increases in turn the potential for hydrophobic interactions, so that urea alone is often not sufficient to quench completely the hydrophobic interactions especially when lipids are present in the sample. This explains why detergents, which can be viewed as specialized agents for hydrophobic interactions, are almost always included in the urea-based solubilization mixtures for 2D electrophoresis. Detergents act on hydrophobic interactions by providing a stable dispersion of a hydrophobic medium in the aqueous medium, through the presence of micelles for example. Therefore, the hydrophobic molecules (e.g. lipids) are no longer collapsed in the aqueous solvent but will disaggregate in the micelles, provided the amount of detergent is sufficient to ensure maximal dispersion of the hydrophobic molecules.

Detergents have polar heads that are able to contract other types of noncovalent bonds (hydrogen bonds, salt bonds for charged heads, etc.). The action of detergents is the sum of the dispersive effect of the micelles on hydrophobic part of the molecules and the effect of their polar heads on the other types of bonds. This explains why various detergents show very variable effects varying from a weak and often incomplete delipidation (e.g. Tweens) to a very aggressive action where the exposure of the hydrophobic core in the detergent-containing solvent is no longer energetically unfavored and leads to denaturation (e.g. SDS).

Of course, detergents used for IEF must bear no net electrical charge, and only nonionic and zwitterionic detergents may be used. However, ionic detergents such as SDS may be used for the initial solubilization, prior to isoelectric focusing, in order to increase solubilization and facilitate the removal of interfering compounds. Low amounts of SDS can be tolerated in the subsequent IEF [10] provided that high concentrations of urea [16] and nonionic [10] or zwitterionic detergents [17] are present to ensure complete removal of the SDS from the proteins during IEF. Higher amounts of SDS must be removed prior to IEF, by precipitation [9] for example. It must therefore be kept in mind that SDS will only be useful for solubilization and for sample entry, but will not cure isoelectric precipitation problems.

The use of nonionic or zwitterionic detergents in the presence of urea presents some problems due to the presence of urea itself. In concentrated urea solutions, urea is not freely dispersed in water but can form organized channels (see [18]). These channels can bind linear alkyl chains, but not branched or cyclic molecules, to form complexes of undefined stoichiometry called inclusion compounds. These complexes are much less soluble than the free solute, so that precipitation is often induced upon formation of the inclusion compounds, precipitation being stronger with increasing alkyl chain length and higher urea concentrations. Consequently, many nonionic or zwitterionic detergents with linear hydrophobic tails [19], [20] and some ionic ones [21] cannot be used in the presence of high concentrations of urea. This limits the choice of detergents mainly to those with nonlinear alkyl tails (e.g. Tritons, Nonidet P40, CHAPS) or with short alkyl tails (e.g. octyl glucoside), which are unfortunately less efficient in quenching hydrophobic interactions. Sulfobetaine detergents with long linear alkyl tails have however received limited applications, as they require low concentrations of urea. Good results have been obtained in certain cases for sparingly soluble proteins [22-24], although this type of protocol seems rather delicate owing to the need for a precise control of all parameters to prevent precipitation.

Apart from the problem of inclusion compounds, the most important problem linked with the use of urea is carbamylation. Urea in water exists in equilibrium with ammonium cyanate, the level of which increases with increasing temperature and pH [25]. Cyanate can react with amines to yield substituted urea. In the case of proteins, this reaction takes place with the a-amino group of the N-terminus and the e-amino groups of lysines. This reaction leads to artefactual charge heterogeneity, N-terminus blocking and adduct formation detectable in mass spectrometry. Carbamylation should therefore be completely avoided. This can be easily made with some simple precautions. The use of a pure grade of urea (p.a.) decreases the amount of cyanate present in the starting material. Avoidance of high temperatures (never heat urea-containing solutions above 37°C) considerably decreases cyanate formation. In the same trend, urea-containing solutions should be stored frozen (-20°C) to limit cyanate accumulation. Last but not least, a cyanate scavenger (primary amine) should be added to urea-containing solutions. In the case of isoelectric focusing, carrier ampholytes are perfectly suited for this task. If these precautions are correctly taken, proteins seem to withstand long exposures to urea without carbamylation [26].

## 3. Solubility During IEF

Additional solubility problems often arise during the IEF at sample entry and solubility at the isoelectric point.

### 3.1. Solubility during sample entry

Sample entry is often quite critical. In most 2D systems, sample entry in the IEF gel corresponds to a transition between a liquid phase (the sample) and a gel phase of higher friction coefficient. This induces a stacking of the proteins at the sample-gel boundary, which results in very high concentration of proteins at the application point. These concentrations may exceed the solubility threshold of some proteins, thereby inducing precipitation and sometimes clogging of the gel, with poor penetration of the bulk of proteins. Such a phenomenon is of course more prominent when high amounts of proteins are loaded onto the IEF gel. The sole simple but highly efficient remedy to this problem is to include the sample in the IEF gel. This process abolishes the liquid-gel transition and decreases the overall protein concentration, as the volume of the IEF gel is generally much higher than the one of the sample.

This process is however rather difficult for tube gels in carrier ampholyte-based IEF. The main difficulty arises from the fact that the thiol compounds used to reduce disulfide bonds during sample preparation are strong inhibitors of acrylamide polymerization, so that conventional samples cannot be used as such. Alkylation of cysteines and of the thiol reagent after reduction could be a solution, but many neutral alkylating agents (e.g. iodoacetamide, N-ethyl maleimide) also inhibit acrylamide polymerization. Owing to this situation, most workers describing inclusion of the sample within the IEF gel have worked with nonreduced samples [27,28]. Although this presence of disulfide bridges is not optimal, inclusion of the sample within the gel has proven of great but neglected interest [27,28]. It must however be pointed out that it is now possible to carry out acrylamide polymerization in an environment where disulfide bridges are reduced. The key is to use 2mM tributylphosphine as the reducing agent in the sample and using tetramethylurea as a carrier solvent. This ensures total reduction of disulfides and is totally compatible with acrylamide polymerization with the standard Temed/persulfate initiator (T. Rabilloud, unpublished results). This modification should help the experimentators trying sample inclusion within the IEF gel when high amounts of proteins are to be separated by 2D.

The process of sample inclusion within the IEF gel is however much simpler for IPG gels. In this case, rehydration of the dried IPG gel in a solution containing the protein sample is quite convenient and efficient, provided that the gel has a sufficiently open structure to be able to absorb proteins efficiently [15]. Coupled with the intrinsic high capacity of IPG gels, this procedure enables to easily separate milligram amounts of protein [15].

3.2. Solubility at the isoelectric point

This is usually the second critical point for IEF. The isoelectric point is the pH of minimal solubility, mainly because the protein molecules have no net electrical charge. This abolishes the electrostatic repulsion between protein molecules, which maximizes in turn protein aggregation and precipitation.

The horizontal comet shapes frequently encountered for major proteins and for sparingly soluble proteins often arise from such a near-isoelectric precipitation. Such isoelectric precipitates are usually easily dissolved by the SDS solution used for the transfer of the IEF gel onto the SDS gel, so that the problem is limited to a loss of resolution, which however precludes the separation of high amounts of proteins.

The problem is however more severe for hydrophobic proteins when an IPG is used. In this case, a strong adsorption of the isoelectric protein to the IPG matrix seems to occur, which is not reversed by incubation of the IPG gel in the SDS solution. The result is severe quantitative losses, which seem to increase with the hydrophobicity of the protein and the amount loaded [29]. The sole solution to this serious problem is to increase the protein solubilizing power of the medium used for IEF, by acting both on the chaotrope and on the detergent.

As to the chaotrope, it has been shown that using a mixture of urea and thiourea increases protein solubility [30]. Thiourea has been shown to be a much stronger denaturant than urea itself [31] on a molar basis. Thiourea alone is weakly soluble in water (ca 1M), so that it cannot be used as the sole chaotrope. However, thiourea is more soluble in concentrated urea solutions [31]. Consequently, urea-thiourea mixtures (typically 2M thiourea and 5 to 8M urea, depending on the detergent used) exhibit a superior solubilizing power and are able to increase dramatically the solubility of membrane or nuclear proteins in IPG gels as well as protein transfer to the second dimension SDS gel [30].

The benefits of using thiourea-urea mixtures to increase protein solubility can be transposed to conventional, carrier ampholyte-based focusing in tube gels with minor adaptations. Thiourea strongly inhibits acrylamide polymerization with the standard temed/persulfate system. However, photopolymerization with methylene blue, sodium toluene sulfinate and diphenyl iodonium chloride [32] enables acrylamide polymerization in the presence of 2M thiourea without any deleterious effect in the subsequent 2D [33] so that higher amounts of proteins can be loaded without loss of resolution [33].

As to the detergent, considerable interest has been put in this field due to its potential application for the solubilization of membrane proteins [34]. It must be kept in mind, however, that the detergents used in denaturing IEF must work in high concentrations of urea. On the one hand, this poses the problem of inclusion compounds, as described above. On the other hand, this highly chaotropic mixture changes dramatically the detergent aggregations parameters (critical micellar concentration, critical micellar temperature) and thus the detergents properties. This can be favorable in some cases, e.g. with deoxychaps which cannot be used in water alone due to its high critical micellar temperature (55°C), while it can be used in 8M urea where it is fully soluble at room temperature [35].

Investigations in the field of detergents for denaturing IEF have concerned the two families that are compatible with IEF, namely zwitterionic detergents and nonionic ones. Sulfobetaines with various hydrophobic parts and/or linkers between the hydrophobic and hydrophilic parts have been synthetized and tested [35] [36], [37]. Some of them have shown interesting solubilizing properties, such as ASB14 and C7BzO, and are now commercially available.

Besides this work on this particular detergent family, there has been a renewal of interest on nonionic detergents, and some of them have been shown to be able to solubilize membrane proteins [38], [39], [40]. From this work , the importance of the detergent/chaotrope couple is clearly highlighted. For example, Triton X100 has been used in conjuction with urea since the very beginning of 2D electrophoresis and has not been shown to solubilize any membrane protein. However, when Triton X100 is used with a urea/thiourea mixture, it has been shown to solubilize some membrane proteins efficiently [38].

4. Concluding Remarks

Although this outline chapter has mainly dealt with the general aspects of solubilization, the main concluding remark is that there is no universal solubilization protocol. Standard urea-reducer-detergent mixtures usually achieve disruption of disulfide bonds and noncovalent interactions. Consequently, the key issues for a correct solubilization is the removal of interfering compounds, blocking of protease action, and disruption of infrequent interactions (e.g. severe ionic bonds). These problems will strongly depend on the type of sample used, the proteins of interest and the amount to be separated, so that the optimal solubilization protocol can vary greatly from a sample to another.

However, the most frequent bottleneck for the efficient 2D separation of as many and as much proteins as possible does not usually lie in the initial solubilization but in keeping the solubility along the IEF step. In this field, the key feature is the disruption of hydrophobic interactions, which are responsible for most, if not all, of the precipitation phenomena encountered during IEF. This means improving solubility during denaturing IEF will focus on the quest of ever more powerful chaotropes and detergents. In this respect, the use of thiourea may prove to be one of the keys to increase the solubility of proteins in 2D electrophoresis. One of the other keys being the use of as powerful detergent or detergent mixtures as possible. Among a complex sample, some proteins may be well denatured and solubilized by a given detergent or chaotrope, while other proteins will require another detergent or chaotrope. Consequently, the future of solubilization may well be to find mixtures of detergents and chaotropes able to cope with the diversity of proteins encountered in the complex samples separated by 2D electrophoresis. It must be kept in mind, however, that this protein diversity may overcome the solubilization power that is achievable with chemicals bearing no electrical charge, as in the case for IEF. When hydrophobic proteins are to be analyzed, it may be a safer approach to use ionic detergents. These have a much higher solubilizing power, as they confer a net electrical charge to the protein-detergent complexes, and the coulombian repulsion between the protein detergent complexes prevents aggregation and promotes solubilization. The price to pay is to renounce to IEF and to use electrophoresis schemes of much lower resolution [41]. However, such electrophoresis schemes using only ionic detergents have been shown to be able to deal with very hydrophobic proteins [42]